\documentclass[11pt]{article}
\usepackage{amsmath,epsfig,subfigure,amsthm,amsfonts,amsbsy,amssymb,latexsym,amsxtra}
\usepackage{empheq}

\usepackage{tikz}
\usetikzlibrary{fit,shapes.misc}

\newcommand\marktopleft[1]{%
    \tikz[overlay,remember picture] 
        \node (marker-#1-a) at (0,1.5ex) {};%
}
\newcommand\markbottomright[1]{%
    \tikz[overlay,remember picture] 
        \node (marker-#1-b) at (0,0) {};%
    \tikz[overlay,remember picture,thick,dashed,inner sep=3pt]
        \node[draw,rounded rectangle,fit=(marker-#1-a.center) (marker-#1-b.center)] {};%
}

\newcommand*\widefbox[1]{\fbox{\hspace{2em}#1\hspace{2em}}}

\newtheorem{thm}{Theorem}

\newtheorem{rem}[thm]{Remark}
\newcommand{\N}{\mathbb{N}}
\newcommand{\R}{\mathbb{R}}
\newcommand{\E}{\mathbb{E}}
\renewcommand{\P}{\mathbb{P}}

\newcommand{\Exp}{\mathbb{E}}
\newcommand{\Pro}{\mathbb{P}}

\newcommand{\df}{\,\mathrm{d}}
\newcommand{\be}{\begin{eqnarray}}
\newcommand{\ee}{\end{eqnarray}}
\newcommand{\by}{\begin{eqnarray*}}
\newcommand{\ey}{\end{eqnarray*}}
\newcommand{\td}{\tilde}

\begin{document}
\title{Conditional Asian Options}
\author{
    Runhuan Feng\\
    Department of Mathematics\\
    University of Illinois at Urbana-Champaign\\
    rfeng@illinois.edu
  \and
    Hans W. Volkmer\\
    Department of Mathematical Sciences\\
    University of Wisconsin - Milwaukee\\
    volkmer@uwm.edu
}

\maketitle

\begin{abstract}

Conditional Asian options are recent market innovations, which offer cheaper and long-dated alternatives to regular Asian options. In contrast with payoffs from regular Asian options which are based on average asset prices, the payoffs from conditional Asian options are determined only by average prices above certain threshold. Due to the limited inclusion of prices, conditional Asian options further reduce the volatility in the payoffs than their regular counterparts and have been promoted in the market as viable hedging and risk management instruments for equity-linked life insurance products. There has been no previous academic literature on this subject and practitioners have only been known to price these products by simulations. We propose the first analytical approach to computing prices and deltas of conditional Asian options in comparison with regular Asian options. In the numerical examples, we put to the test some cost-benefit claims by practitioners. As a by-product, the work also presents some distributional properties of the occupation time and the time-integral of geometric Brownian motion during the occupation time.


\end{abstract}

\section{Introduction}

Asian options (average options) are widely used in commodity and stock markets as cheaper alternatives to European and American options for hedging and risk management. A common use of Asian options is to transfer the risk of volatility in asset prices to capital markets. Consider a continuous-time stochastic model for asset prices, denoted by $\{X_t, t\ge 0\}$ with $X_0=x$. The payoff from a continuous European-style Asian put option with fixed strike $K$ and maturity $T$ is determined by
\be \left(K- \frac1T \int^T_0 X_t \df t \right)_+ ,\label{reg} \ee where $(x)_+=\max\{x,0\}$. A review of historical references and examples of Asian options can be found in \cite{Lin04b}. The pricing of Asian options led to a flourishing of computational techniques in the mathematical finance literature. Monte Carlo simulations were among the first pricing methods, proposed by \cite{KemVor}, \cite{BoyBroGla}. Bounds and approximations based on conditional expectations can be seen in \cite{Cur}, \cite{NieSan}, \cite{Lor}.  Numerical PDE methods were extensively studied in a series of papers including \cite{RogShi}, \cite{AlzDecKoe},  \cite{Vec}, \cite{ZvaVetFor}. Numerical inversion of Laplace/Fourier transforms were developed in \cite{GemYor}, \cite{CarCle}, \cite{FuMadWan}, \cite{CaiKou},  \cite{CarSch}. Latest work by \cite{CheFenLin} also extends Hilbert transform methods to Asian option pricing. A variety of orthogonal polynomial/eigenfunction expansions were used in \cite{Duf}, \cite{Lin04b}, \cite{Ju}, and \cite{ZhaOos}. Asymptotics expansions of Asian option prices were developed in \cite{CarPetYor}; \cite{CaiLiShi}, \cite{Ger}, etc. Given the sheer volume of research papers on this subject matter, the list of aforementioned papers is by no means a comprehensive review of the literature.

\begin{figure}[h]
\begin{center}
\includegraphics[height=7cm,width=10cm]{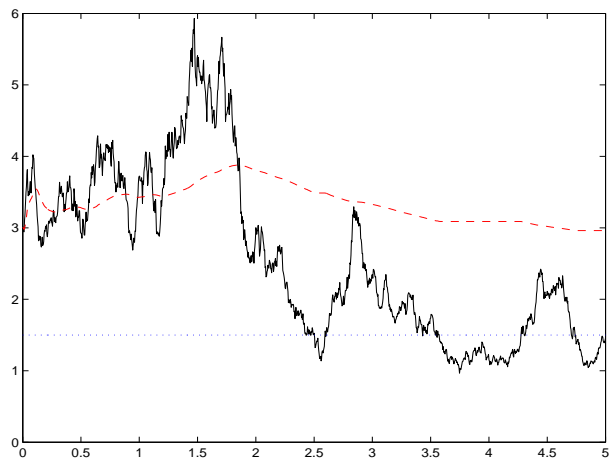}
\end{center}
\caption{Example of average-to-date above a threshold. }
\label{fig:sample}
\end{figure}

In recent years, BNP Paribas introduced a variation of the Asian option, termed conditional Asian option, under which the average of asset prices is only based on prices that are above a certain threshold (See an industrial presentation by \cite{Seg}).  Formulated mathematically, the payoff from the continuous conditional Asian option is given by
\be \left(K- \frac{\int^T_0 X_t I_{\{X_t> b\}} \df t}{\int^T_0 I_{\{X_t> b\}} \df t} \right)_+ ,\label{cond}\ee where the threshold level is denoted by $b>0$ and $I_A$ is an indicator function which equals $1$ if $A$ is true and $0$ otherwise. Figure \ref{fig:sample} gives a sample path of asset prices (black solid line) where the average-to-date (red dash line) only includes prices above a threshold (blue dotted line). Observe that the sample path of the average-to-date becomes flat whenever asset prices plunge below the threshold. Parameters used in this numerical example are provided in Section \ref{sec:num}.
 In the BNP Paribas presentation, the threshold $b$ is called observation barrier and is $50\%$ of the initial price $x_0$. They gave an illustrative example of a five-year at-the-money conditional Asian option ($K=X_0$), which is the discrete version of  \eqref{cond} with monthly observations.

There are several reasons why conditional Asian puts may be considered effective alternatives to regular Asian puts in the financial market. First, conditional Asian puts are cheaper than Asian puts with the same term and strike. Observe that
\by \int^T_0 \frac{X_t I_{\{X_t>b\}}}{\int^T_0 I_{\{X_t>b\}} \df t } \df t
&=&\frac1T\int^T_0 X_t I_{\{X_t>b\}} \df t+\frac1T \int^T_0 \frac{\int^T_0 I_{\{X_t\le b\}} \df t}{\int^T_0 I_{\{X_t>b\}} \df t} X_t I_{\{X_t>b\}} \df t\\
&\ge& \frac1T\int^T_0 X_t I_{\{X_t>b\}} \df t+ \frac1T \int^T_0  b I_{\{X_t\le b\}} \df t\\
&\ge& \frac1T\int^T_0 X_t I_{\{X_t>b\}} \df t+  \frac{ 1}{T}\int^T_0  X_t I_{\{X_t\le b\}} \df t=\frac{ 1}{T}\int^T_0  X_t \df t
.\ey Thus the payoff from a conditional Asian option \eqref{cond} is surely no greater than that from a regular Asian option \eqref{reg}.
According to BNP Paribas, a five-year at-the-money conditional Asian option was shown to cost $40\%$ less in comparison with a regular Asian option, yet achieved $75\%$ of the initial delta for a regular one. Second, most Asian options available in the market are relatively short termed (typically less than one year). In fact, there are very few hedging products that can match the duration of insurance liabilities due to their long-term nature. BNP Paribas promoted five-year conditional Asian options as hedging solutions for variable annuities, which are investment-combined insurance products with very long term (typically more than ten years). Third, cash flow structures of variable annuity guaranteed benefits resemble that of Asian options. This observation was exploited in \cite{FenVol, FenVol2, Feng} for the computation of risk measures for guaranteed benefits. It follows from the same logic that long-term Asian-like options provide natural hedge against financial risks embedded in variable annuity products. Therefore, conditional Asian options are touted as a cost effective tool for asset and liability management.

In this paper, we intend to develop an analytical method to price the conditional Asian option and compute its delta. We consider the Black-Scholes model under which the underlying asset follows the geometric Brownian motion under the risk-neutral measure
\by \df X_t= r X_t \df t +\sigma X_t \df B_t, \ey and under which the no-arbitrage price of a conditional Asian put option can be written as
\by
\mathrm{AP}_b&:=&\td{\Exp}\left[ e^{-rT}\left(K- \frac{\int^T_0 X_t I_{\{X_t> b\}} \df t}{\int^T_0 I_{\{X_t> b\}} \df t} \right)_+ \right],
\ey where $\td{\Exp}$ denotes the expectation under the risk-neutral probability measure. The regular Asian put is clearly a special case of conditional Asian put with $b=0$ and hence its price is given by $\mathrm{AP}_0$.
It turns out that the price of conditional Asian put is best computed as the price of regular Asian, which can be done through inversion of a closed-form Laplace transform, minus some price spread, which needs to be worked out by inversion of a Laplace transform with an integral representation.
\[\mathrm{AP}_b=\mathrm{AP}_0-\mbox{Price Spread}.\]
The price spread is clearly a result of cost saving from filtering out undesirable low prices in the average price. A practical advantage of such a split is that one can compute the price spread directly without having to carry out simulations for pricing conditional Asian options. Practitioners may use it to extrapolate conditional Asian prices by combining market prices of Asian options with estimated price spread. The deterministic algorithm developed later for the computation of price spread can be easily used with implied volatility, making the conditional Asian option pricing consistent with the standard practice of pricing option with less liquidity and thin trading.

\section{Asian options revisited} \label{sec:asian}

As it shall become clear later, the regular Asian option plays an important role in the pricing of conditional Asian options. We shall present some results on Asian put option prices, which can be shown to be consistent to known results on Asian call option prices in the literature.

Consider the process
\be Y_t:=\int^t_0 X_s \df s,\qquad t \ge 0. \label{Y}\ee Let $p(x,t,y)$ be the transition density of $Y$ with $X_0=x$ and set
\be  P(x,t,y)=\int_0^y p(x,t,v)\,dv, \quad Q(x,t,y)=\int_0^y P(x,t,v)\,dv .\label{defPQ}\ee
Then it follows immediately that
\be \boxed{\mathrm{AP}_0=\frac1T e^{-rT} Q(x,T, TK),}\label{ap0}\ee
In what follows, we derive an explicit solution to the Laplace transform of $Q(x,t,y)$ with respect to $t$ only. This allows computation of $Q$ by a one-dimensional numerical Laplace inversion.

In \cite{FenVol2} we considered the special case
\[ \mathfrak{X}_t=\exp(2\nu t+2 B_t),\qquad \mathfrak{Y}_t=\int^t_0 \mathfrak{X}_s \df s.\]
Setting $u=4w/\sigma^2$ and using the scaling property $B_{ct}\sim \sqrt{c} B_t$, we get
\begin{eqnarray*}
 Y_t=\int_0^t X_u\,du&\sim& x \frac{4}{\sigma^2} \int_0^{\frac{\sigma^2}{4} t} \exp(2\nu w+2 B_w)\,dw=\frac{4x}{\sigma^2}\mathfrak{Y}_{\sigma^2 t/4} ,
\end{eqnarray*}
where the constant $\nu$ is given by \[ \nu=\frac{2r}{\sigma^2} -1 .\]
We denote the transition density of $\mathfrak{Y}$ by $p_0$ and define $P_0, Q_0$ in the same way in which $P, Q$ are defined by $p$ in \eqref{defPQ}. It follows immediately that
\begin{eqnarray}
 &&P(x,t,y)=P_0\left(\frac{\sigma^2 t}{4}, \frac{\sigma^2 y}{4 x}\right) \label{eq:P} \\
  &&p(x,t,y)= \frac{\sigma^2}{4x} p_0\left(\frac{\sigma^2 t}{4} \frac{\sigma^2 y}{4x}\right) \nonumber\\
 &&\boxed{Q(x,t,y)= \frac{4x}{\sigma^2} Q_0\left(\frac{\sigma^2 t}{4} , \frac{\sigma^2 y}{4 x}\right)  .} \label{eq:Q}
\end{eqnarray}
Therefore, the remaining task is to find a way of computing $Q_0$.

We shall denote the Laplace transform $\td{f}(s):=\int^\infty_0 e^{-st} f(t) \df t$.
It is known that
\be \tilde{p}_0(s,y):=\int_0^\infty e^{-st}p_0(t,y)\,dt =\frac{\Gamma(\eta-\kappa+\frac12)}{\Gamma(1+2\eta)}2^{-\kappa} f_\kappa(y), \label{p0}
\ee
where
\[ f_\kappa(y)=y^{-\kappa}\exp\left(-\frac{1}{4y}\right)M_{\kappa,\eta}\left(\frac1{2y}\right), \]
with $M_{\kappa,\eta}$ denoting a Whittaker function, and
\[  \kappa=\frac{1-\nu}{2}, \quad \eta=\frac12 \sqrt{2s+\nu^2}.\]
We use integration by parts to show that
\[ \int_0^y f_\kappa(z)\,dz = \frac{1}{\eta+\kappa-\frac12}\left(2^{\kappa-1} \frac{\Gamma(1+2\eta)}{\Gamma(\eta-\kappa+\frac32)}-f_{\kappa-1}(y)\right) .\]
It follows that
\be \tilde P_0(s,y)=\frac{1}{2(\eta+\kappa-\frac12)(\eta-\kappa+\frac12)}-\frac{\Gamma(\eta-\kappa+\frac12)}{(\eta+\kappa-\frac12)\Gamma(1+2\eta)}2^{-\kappa}f_{\kappa-1}(y) \label{eq:tP0}\ee
and
\begin{empheq}[box=\widefbox]{align}
 \tilde Q_0(s,y)&=\frac{y}{2(\eta+\kappa-\frac12)(\eta-\kappa+\frac12)}-\frac{1}{4(\eta+\kappa-\frac12)(\eta+\kappa-\frac32)(\eta-\kappa+\frac32)(\eta-\kappa+\frac12)} \nonumber
\\ &+\frac{\Gamma(\eta-\kappa+\frac12)}{\Gamma(1+2\eta)}\frac{2^{-\kappa}}{(\eta+\kappa-\frac12)(\eta+\kappa-\frac32)} f_{\kappa-2}(y) . \label{eq:Q0}
\end{empheq}

Although the expression for $\tilde{Q}_0$ was not explicitly stated in the literature, similar expressions are known for Asian call options. \cite[(3.10)]{GemYor} developed a single integral expression for a similar Laplace transform of Asian call price with respect to the time parameter, which was later improved by \cite{DonGhoYor} in terms of Kummer's function of the first kind and similarly by \cite[Section 3]{Sha}. It can be shown through the put-call parity that the result here is consistent with that of Asian call price in \cite[(3.8)]{DonGhoYor}. In the first example of Section \ref{sec:num} we shall verify numerically the equivalence of the formula in this section and those known in the literature.

\section{Joint Laplace transform}

The computation of the conditional Asian option price requires the study of the joint distribution of the pair $(U_t, V_t)$ where
\[ U_t=\int_0^t I_{\{X_\tau>b\}}\,d\tau,\quad V_t= \int_0^t X_\tau I_{\{X_\tau>b\}}\,d\tau,\qquad b\ge 0.\]
The former is known as an occupation time whose transition density is studied in \cite{Pec}. The latter is known in the literature only in the case of $b=0$, which is also called Yor's process.

Let us consider the triple joint Laplace transform
\be F(b,x,s,\alpha,\beta)=\frac1 s \tilde{\Exp}^x [\exp\{-\alpha U_{T_s} -\beta V_{T_s}\}], \label{trilap} \ee
where $T_s$ is an exponential random variable with mean $1/s$ independent of $X$.
By the Feynman-Kac theorem, $F(x)=F(b,x,s,\alpha,\beta)$ satisfies the
differential equation
\begin{equation}\label{3:ode}
\tfrac12\sigma^2 x^2 F''(x)+rxF'(x)-\left[(\alpha+\beta x)I_{\{x>b\}}+s\right] F(x)=-1 ,\quad x>0
\end{equation}
with boundary conditions
\begin{eqnarray}
\lim_{x\to0+} F(x)&=&\frac1s,\nonumber \\
\lim_{x\to\infty} F(x)&=&0 .\label{2:bc1}
\end{eqnarray}
As usual, a solution $F(x)$ of \eqref{3:ode} is understood to be differentiable at $x=b$ with a jump in the second derivative there.

\subsection{The case $b=0$}

We first consider the simpler case where $b=0$. In this case, $U_t=t$ and
\[F(0,x,s,\alpha,\beta)=\frac1 s \tilde{\Exp}^x [\exp\{-\alpha T_s -\beta V_{T_s}\}].\]
Then $F(x)=F(0,x,s,\alpha,\beta)$ satisfies the ODE
\begin{equation}\label{2:ode}
 \tfrac12\sigma^2 x^2 F''(x)+rxF'(x)-(\alpha+s+\beta x)F(x)=-1,\qquad x >0,
\end{equation} subject to the boundary conditions \eqref{2:bc1} and
\be \lim_{x\to0+} F(x)&=&\frac1{s+\alpha}. \label{2:bc2}\ee
The homogeneous equation
\begin{equation}\label{2:ode2}
 \tfrac12\sigma^2 x^2 F''(x)+rxF'(x)-(\alpha+s+\beta x)F(x)=0
\end{equation}
has the fundamental system of solutions
\begin{eqnarray}
 F_1(x)&=&x^{-(1+\mu)/2}I_{\lambda}\left(\frac{2}{\sigma}\sqrt{2\beta x}\right),\nonumber \\
 F_2(x)&=&x^{-(1+\mu)/2}K_{\lambda}\left(\frac{2}{\sigma}\sqrt{2\beta x}\right) , \label{F2}
\end{eqnarray}
where $I_\lambda, K_\lambda$ denote modified Bessel functions and
\begin{eqnarray*}
 \mu&=& \frac{2r}{\sigma^2}-2,\\
 \lambda&=& \sqrt{(\mu+1)^2+8\sigma^{-2}(s+\alpha)}.
\end{eqnarray*}
When $\alpha,\beta,s>0,$ the solution $F_1(x)$ is positive and increasing on $(0,\infty)$ with $\lim_{x\to0+} F_1(x)=0$, $\lim_{x\to\infty}F_1(x)=\infty$
while $F_2(x)$ is positive and decreasing on $(0,\infty)$ with $\lim_{x\to0+} F_2(x)=\infty$, $\lim_{x\to\infty}F_2(x)=0$.
The Wronskian is
\[ F_1(x)F_2'(x)-F_1'(x)F_2(x)=-\tfrac12 x^{-2-\mu}. \]
From the variation of parameters formula, we obtain
\begin{equation}\label{2:FF}
F(x)=\frac{4}{\sigma^2} F_1(x)\int_x^\infty z^\mu F_2(z)\,dz +\frac{4}{\sigma^2}F_2(x)\int_0^x z^\mu F_1(z)\,dz.
\end{equation}
Note that the integrals in \eqref{2:FF} are well-defined.

Certainly $F(x)$ defined by \eqref{2:FF} satisfies \eqref{2:ode}.
To verify \eqref{2:bc1} we use the known behavior of $I_\lambda(z)$, $K_\lambda(z)$ as $z\to0$:
\begin{eqnarray}\label{2:IK0}
I_\lambda(z)&\sim&\frac{\left(\frac{z}{2}\right)^\lambda}{\Gamma(\lambda+1)},\label{2:I0}\\
K_\lambda(z)&\sim & \frac12 \Gamma(\lambda)\left(\frac{z}{2}\right)^{-\lambda}.\label{2:K0}
\end{eqnarray}
Then, by using L'Hospital's rule, we get
\[ \lim_{x\to0+} F(x)=\frac{4}{\sigma^2} \frac{1}{\lambda(\lambda-\mu-1)}+\frac{4}{\sigma^2} \frac{1}{\lambda(\lambda+\mu+1)}=\frac1{s+\alpha}.\]
In a similar way, we can show that also \eqref{2:bc2} is true.
Note that since $\lim_{x\to\infty}F_1(x)=\infty$ and $\lim_{x\to0+} F_2(x)=\infty$, there is only one solution of \eqref{2:ode}, \eqref{2:bc1}, \eqref{2:bc2}, so $F(x)$ must be the
function we are looking for.

In the Appendix we use the theory of Lommel's functions to show that $F$ in \eqref{2:FF} can be written in the form
\begin{empheq}[box=\widefbox]{align}
F(0,x,s,\alpha,\beta)&=\frac{1}{s+\alpha}{}_1F_2(1;\tfrac12(\mu-\lambda+3),\tfrac12(\mu+\lambda+3);2\sigma^{-2}\beta x)\nonumber\\
&+\frac{2}{\sigma^2}\Gamma(\tfrac12(\mu-\lambda+1))\Gamma(\tfrac12(\mu+\lambda+1)) \left(\frac{1}{\sigma}\sqrt{2\beta}\right)^{-1-\mu}F_1(x),\label{2:FFF}
\end{empheq}
where ${}_1F_{2}$ is a generalized hypergeometric function defined for all $z\in \mathbb{C}$ by the convergent series
\[_1F_2(a_1;b_1,b_2;z)=\sum^\infty_{k=0} \frac{(a_1)_k}{(b_1)_k (b_2)_k}\frac{z^k}{k!},\qquad (b_1, b_2 \ne 0,-1,-2,\cdots)\] with the Pochhammer's symbol $(\alpha)_n=\alpha(\alpha+1)\cdots(\alpha+n-1)$ if  $n=1,2,\cdots$ and $(\alpha)_0=1$.

The advantage of \eqref{2:FFF} over \eqref{2:FF} is that it does not involve an integration. The disadvantage is that \eqref{2:FFF} breaks down when $\lambda-\mu=2p+1$, $p=1,2,\dots$
The condition $\lambda-\mu=2p+1$ is equivalent to
\[ p\frac{2r}{\sigma^2}-\frac{2(s+\alpha)}{\sigma^2}+p(p-1)=0.\]
However, we will use \eqref{2:FFF} only for purely imaginary values of $\alpha,\beta$ and then the exceptional cases do not occur.
Another problem with \eqref{2:FFF} is that we might have huge cancellations in the addition of the two summands. Using arbitrary precision software the problem can be solved by increasing the number of digits although this is not an elegant method. Later we will make sure that we use \eqref{2:FFF} only in cases when there are no large cancellations.

\subsection{The case $b>0$}

If $x>b$ then there is a constant $A$ such that
\[ F(x)=A F_2(x)+ Y(x),\]
where $F_2$ is defined as in the previous subsection and $Y$ is given by \eqref{2:FF} or \eqref{2:FFF}, i.e. \[\boxed{Y(x)=F(0,x,s,\alpha,\beta).}\]
Note that we could choose $Y(x)$ differently, but it must be a solution of \eqref{3:ode} which converges to $0$ as $x\to\infty$.
If $0<x<b$, then there is a constant $B$ such that
\[ F(x)=B x^\rho+\frac{1}{s} ,\]
where
\by  \rho&=&-\tfrac12(\mu+1)+\tfrac12 \lambda_0,\\
\lambda_0&=&\sqrt{(\mu+1)^2+8\sigma^{-2}s}. \ey
By matching the values of $F$ and $F'$ at $x=b$ we obtain for $x>b$
\begin{equation}\label{3:F}
\boxed{F(b,x,s,\alpha,\beta)=\Phi(b,x,s,\alpha,\beta)+Y(x),}
\end{equation}
where
\begin{equation}\label{3:Phi}
\boxed{\Phi(b,x,s,\alpha,\beta)=\frac{\rho\left(\frac1s-Y(b)\right)+bY'(b)}{\rho F_2(b)-bF_2'(b)} F_2(x).}
\end{equation}
Observe that as $b\rightarrow 0$, $\Phi(b,x,s,\alpha,\beta)$ approaches zero due to \eqref{2:bc1}.

\section{Moments of $(U_t, V_t)$} \label{sec:moment}

We observe that when $\alpha=i\tau z$ and $\beta=-i\tau,$ the mapping $\tau \to sF(b,x,s,\alpha,\beta)$ becomes the characteristic function of $W:=V_{T_s}-zU_{T_s},$ i.e.
\[sF(b,x,s,i\tau z,-i\tau)=\tilde{\Exp}^x[\exp\{i\tau W\}].\]
In this section we will determine the coefficients $E_k$ in the expansion 
\begin{equation}\label{psiasy}
 sF(b,x,s,i\tau z,-i\tau)=E_0+E_1\tau+\dots+E_p\tau^p+o(\tau^p)\quad \text{as $\tau\to 0$.}
\end{equation}
Then the moments of $W$ are given by 
\[ \tilde{\Exp}^x[W^k]= i^{-k} k! E_k;\] 
see \cite[Theorem 2.3.3]{Lukacs}.
Not only does \eqref{psiasy} lead to a way of computing moments of $(U_t, V_t)$, it also shows the rate of convergence near $\tau=0$ of an important integral of $F$ to be seen in the computation of conditional Asian option prices.

Consider the case where $x>b$. Let $p\in\N$. We assume that $\rho>p$ which is equivalent to
\[ \frac{2s}{\sigma^2}> p(p-1)+p\frac{2r}{\sigma^2} .\]
For example, if $p=1$ then we require that $s>r$.
We set $\alpha=i\tau z, \beta=-i\tau$ in \eqref{2:FFF} and use \eqref{2:I0}.
Since $\rho>p$ we see that the second summand on the right-hand side of \eqref{2:FFF} is $o(\tau^p)$ as $\tau\to 0$.
The first term on the right-hand side of \eqref{2:FFF} is analytic at $\tau=0$. Therefore, we can write
\[ sY(x,s,i\tau z,-i\tau)=A_0+A_1\tau+\dots+A_p\tau^p +o(\tau^p) ,\]
where $A_0=1$ and
\[A_1= -i\left(\frac{x}{r-s}+\frac{z}{s}\right) .\]
In a similar way we get ($'=\df /\df x$)
\[ sY'(x,s,i\tau z,-i\tau)= B_1\tau+B_2\tau^2+\dots+B_p\tau^p+o(\tau^p) ,\]
where
\[ B_1=-i\frac{1}{r-s}.\]
It follows that
\[ \rho(1-sY(b)) +b s Y'(b)=C_1\tau+\dots+C_p\tau^p+o(\tau^p),\]
where
\[ C_1=i\rho\left(\frac{b}{r-s}+\frac{z}{s}\right)- i \frac{b}{r-s} .\]
Now consider the function (recall $\alpha=i \tau z, \beta=-i\tau$ )
\begin{equation}\label{omega}
\omega(\tau)= \frac{F_2(x)}{\rho F_2(b)-bF_2'(b)} .
\end{equation}
We express $\omega$ in terms of the Bessel function $I_\lambda$ and $I_{-\lambda}$ by using \eqref{F2},
\[ K_v(u)=\frac{\pi}{2\sin(\pi v)}(I_{-v}(u)-I_v(u)) \]
and
\[ I_v(u)=\left(\frac{u}{2}\right)^v \sum_{k=0}^\infty \frac{u^{2k}}{4^{k}k!\Gamma(v+k+1)}. \]
Upon substitution of $K_\lambda$ in \eqref{F2}, we observe that the term involving $I_{\lambda}$ is $O(\beta^{\lambda/2})$ as $\tau\rightarrow 0$. Thus,
\[F_2(x)=\frac{\pi}{2\sin(\lambda\pi)} x^{-(1+\mu)/2}I_{-\lambda}\left(\frac2\sigma\sqrt{2\beta x}\right) +O\left(\beta^{\lambda/2}\right).\]
We treat the term $\rho F_2(b)-b F_2'(b)$ similarly:
\begin{eqnarray*}
&& \rho F_2(b)-bF_2'(b)= \frac{\lambda_0-\lambda}{2}b^{-(1+\mu)/2}K_\lambda\left(\frac2\sigma\sqrt{2\beta b}\right)+\frac{\sqrt{2\beta}}{\sigma}b^{-\mu/2} K_{\lambda+1}\left(\frac2\sigma\sqrt{2\beta b}\right)\\
&&=\frac{\pi}{2\sin(\lambda \pi)} \frac{\lambda_0-\lambda}{2} b^{-(1+\mu)/2}I_{-\lambda}\left(\frac2\sigma\sqrt{2\beta b}\right)- \frac{\pi}{2\sin(\lambda \pi)} \frac{\sqrt{2\beta}}{\sigma} b^{-\mu/2} I_{-\lambda-1}\left(\frac2\sigma\sqrt{2\beta b}\right) +O\left(\beta^{\lambda/2}\right).
\end{eqnarray*}
Multiplying the numerator and the denominator of the fraction in \eqref{omega} by $\frac{2}{\pi}\sin(\pi\lambda)\left(\frac{\sqrt{2\beta}}{\sigma}\right)^\lambda\Gamma(-\lambda)$ gives
\begin{equation}
\omega(\tau)=\left(\frac{x}{b}\right)^{-(1+\mu+\lambda)/2}\sum_{k=0}^\infty \frac{\left(\frac{8}{\sigma^2} \beta x\right)^k}{4^k k!(-\lambda)_{k+1}}\left[ \sum_{k=0}^\infty \left(\frac{\lambda_0+\lambda}{2}-k\right)  \frac{\left(\frac{8}{\sigma^2} \beta b\right)^k}{4^k k!(-\lambda)_{k+1}}\right]^{-1}+O\left(\tau^p\right),
\end{equation}
where we used that $\lambda_0+\mu+1>0$ implies $\lambda_0>\rho>p$. Expanding the right-hand side in powers of $\tau$ yields
\[ \omega(\tau)=D_0+D_1\tau+\dots+D_p\tau^p+o(\tau^p) ,\]
where
\[  D_0 = \frac{1}{\lambda_0}\left(\frac{x}{b}\right)^{-\frac12(\mu+\lambda_0+1)}.\]
When we put everything together, we obtain \eqref{psiasy}
with $E_0=1$ and
\[ E_1=i\left(\left(\frac{b}{r-s}+\frac{z}{s}\right)\rho -\frac{b}{r-s}\right)\frac{1}{\lambda_0} \left(\frac{x}{b}\right)^{-\frac12(\mu+\lambda_0+1)}
- i\left(\frac{x}{r-s}+\frac{z}{s}\right).\]

Similarly, one can also work out the expansion when $0<x<b$ and
\[E_1=-i\left( \left(\frac{b}{r-s}+\frac{z}{s}\right)\left(\frac{\mu+\lambda_0+1}2\right)+\frac{b}{r-s}\right) \frac1{\lambda_0} \left(\frac{x}{b}\right)^{\rho}.\]
Therefore,
\begin{equation}\label{limit1}
 \lim_{\tau\to 0} \frac{s}{\tau}\Im F(b,x,s,i\tau z,-i\tau) =\frac{E_1}{i}.
\end{equation} 
By identifying the coefficients of $z$'s, we obtain that when $x\ge b,$
\by \displaystyle
\Exp(V_{T_s})&=&\frac{b(\rho-1)}{(r-s)\lambda_0} \left(\frac{x}{b}\right)^{-\frac12(\mu+\lambda_0+1)}-\frac{x}{r-s};\\
\Exp(U_{T_s})&=&\frac{1}{s}-\frac{\rho}{\lambda_0 s} \left(\frac{x}{b}\right)^{-\frac12(\mu+\lambda_0+1)},
\ey and when $0<x<b,$
\by \displaystyle
\Exp(V_{T_s})&=&\frac{b(\mu+\lambda_0-3)}{2(s-r)\lambda_0} \left(\frac{x}{b}\right)^{\rho};\\
\Exp(U_{T_s})&=&\frac{\mu+\lambda_0+1}{2\lambda_0 s} \left(\frac{x}{b}\right)^{\rho}.
\ey

The above method can be used to evaluate the coefficients $E_k$ for $k\ge 2$ with any computer algebra system, such as Maple and Mathematica, and hence higher moments and cross-moments of $(U_{T_s}, V_{T_s})$. We have developed a computer routine to compute these coefficients, which shall be made available on the authors' webpages.

\section{Conditional Asian option}

The key element of pricing conditional Asian option is to find the distribution of the ratio
\be Z_t:=\frac{V_t}{U_t},\qquad t \ge 0.\label{Z}\ee
We denote its cumulative distribution function by
\[  G(b,x,z,t)=\P(Z_t\le z)\]
Since $V_t/U_t >b,$ we must have
\begin{equation}\label{1:G=0}
G(b,x,z,t)=0\quad\text{for $z\le b$}.
\end{equation}

To price a conditional Asian option with the payoff \eqref{cond},
we need to calculate
\begin{equation}\label{1:price}
 \boxed{\mbox{AP}_b=e^{-rT}\E( (K-Z_T)_+) =e^{-rT}\int_0^K G(b,x,z,T)\,dz,}
\end{equation}
which can be shown through integration by parts.

Applying Gurland's formula \cite[Theorem 1]{Gur} to the ratio of random variables $U_{T_s}/V_{T_s}$, we obtain for $s>0$,
\[ s\int^\infty_0 e^{-st} G(b,x,z,t) \df t=\frac12-\frac1{2 \pi i} \mbox{P.V.} \int^\infty_{-\infty} \frac1\tau  s F(b,x,s,i\tau z,-i\tau) \df \tau.\]
By the reflection principle, we must have $F(b,x,s,\overline{\alpha},\overline{\beta})=\overline{F(b,x,s,\alpha,\beta)}$ and thus
\begin{equation}\label{1:Gurland}
 \boxed{\int_0^\infty e^{-st}G(b,x,z,t)\,dt = \frac1{2s}-\frac1\pi\int_0^\infty \frac1\tau \Im F(b,x,s,i\tau z,-i\tau)\,d\tau.}
\end{equation}

As it is difficult to develop a closed-form solution to $G$ itself, we want to carry out the computation of the conditional Asian option price in \eqref{1:price} in three steps. First, we compute the integral on the right-hand side of \eqref{1:Gurland} numerically. In the previous section, we discussed
the behavior of the integrand as $\tau\to0$ and we shall analyze its behavior as $\tau\to+\infty$. Then, we will use the Gaver-Stehfest algorithm for numerical inversion of Laplace transform to obtain $G$. Finally, we perform the numerical integration in \eqref{1:price}. It is interesting to note that the three operations
\begin{itemize}
\item
 Gurland integral \eqref{1:Gurland},
\item
inverse Laplace transform $s\mapsto t$,
\item
integration of distribution function \eqref{1:price}
\end{itemize}
could be carried out in any order.

\begin{rem}
It is worthwhile to mention that the computation of the regular Asian option price using \cite{Yor92}'s integral representation of the joint density of the geometric Brownian motion and its time integral also requires the evaluation of a triple integral. Such computation was demonstrated in \cite{Ish}. Even though the payoff of conditional Asian option is considerably more evolved than its regular counterpart, the complexity of computation appears to be comparable.
\end{rem}

\begin{rem}
In the case of $b=0$ (i.e. regular Asian option), the computation of the Gurland integral \eqref{1:Gurland} was avoided in Section \ref{sec:asian}, because we exploited the fact that  the two-dimensional random vector $(U_t, V_t)$ degenerates to the vector $(t,Y_t)$ driven by a single random variable and hence
\be &&G(0,x,z,T)=P(x,T,Tz),\label{GP}\\
  &&\mathrm{AP}_0=e^{-rT}\int^K_0 G(0,x,z,T) \df z=\frac1{T}e^{-rT} Q(x,T,TK),\nonumber\ee where $Q$ can be obtained numerically by inverting its Laplace transform determined by \eqref{eq:Q} and \eqref{eq:Q0}.
\end{rem}

As we shall show in the next section, the integral on the right-hand side of \eqref{1:Gurland} decays in the order of $O(1/\tau^2)$, making the numerical integration rather slow.
Therefore, we propose a modification to improve the efficiency of computation, which turns out to have an economic interpretation as well.
We introduce
\[D(b,x,z,t)=G(0,x,z,t)-G(b,x,z,t).
\]
It is easy to show that the function
\[ b\to G(b,x,z,t) \]
is decreasing so $D(b,x,z,t)>0$.  Therefore, the price of a conditional Asian option is broke down into the price of a regular Asian option and some price spread,
\be
\boxed{\mathrm{AP}_b=\mathrm{AP}_0+e^{-rT}\int^K_0 D(b,x,z,T) \df z.} \label{APspread}
\ee
Clearly, the regular Asian option price $\mathrm{AP}_0$ can be determined efficiently by methods in Section \ref{sec:asian}.
For the computation of $D$ we use
\begin{equation}\label{1:Gurland2}
\boxed{\int_0^\infty e^{-st}D(b,x,z,t)\,dt=\frac1\pi\int_0^\infty \frac1\tau\Im \Phi(b,x,s,i\tau z,-i\tau)\,d\tau,}
\end{equation}
where
\begin{equation}\label{1:Phi}
\Phi(b,x,s,\alpha,\beta):=F(b,x,s,\alpha,\beta)-F(0,x,s,\alpha,\beta),
\end{equation} which determines the explicit formula for $\Phi$ in \eqref{3:Phi}.
An advantage of working with $D$ is that the integrand of \eqref{1:Gurland2} decays exponentially as $\tau\to\infty$; see Section 6 for details.

\section{Asymptotics as $\tau\to\infty$} \label{sec:asyminfty}

The function $Y(x)=Y(x,s,i\tau z,-i\tau)$ satisfies the differential equation
\begin{equation}\label{inf:ode}
 \tfrac12\sigma^2 x^2Y''(x)+rxY'(x)-(s+i\tau z-i\tau x)Y(x)=-1 .
\end{equation}
To get an approximation of $Y(x)$ as $\tau\to\infty$, we neglect all terms on the left-hand side of \eqref{inf:ode} which do not depend on $\tau$.
We obtain
\begin{equation}\label{inf:Y}
 Y(x,s,i\tau z,-i\tau)\sim \frac{1}{i(z-x)}\frac{1}{\tau} \quad \text{as $\tau\to+\infty$}
\end{equation}
and
\begin{equation}\label{inf:Yp}
\frac{d}{dx} Y(x,s,i\tau z,-i\tau)\sim \frac{1}{i(z-x)^2}\frac1\tau  \quad \text{as $\tau\to+\infty$} .
\end{equation}
The asymptotics can be proved using \cite[Section 10.10]{Olv}. Numerical tests also confirm \eqref{inf:Y}.
We can even get the complete asymptotic expansion
\[ Y(x,s,i\tau z,-i\tau) \sim \sum_{n=0}^\infty \frac{A_n(x)}{\tau^{n+1}}, \]
where $A_0=\frac{1}{i(z-x)}$ as given above and then recursively
\[ \tfrac12 \sigma^2 x^2A_n''+rxA_n'-sA_n-i(z-x)A_{n+1} =0 .\]
Note that \eqref{inf:Y} breaks down when $z=x$ (a turning point appears.)

With the insight of the asymptotics of $Y$, we find two numerical issues with the computation of the Gurland integral \eqref{1:Gurland}.

\begin{enumerate}
	\item Equation \eqref{1:Gurland} contains the integral
\begin{equation}\label{inf:Gurland}
\int_0^\infty \frac{1}{\tau}\Im Y(x,s,i\tau z,-i\tau)\,d\tau .
\end{equation}
It follows from \eqref{inf:Y} that the integrand in \eqref{inf:Gurland}
behaves like a constant times $1/\tau^2$ as $\tau\to \infty$. This decay rate is fast enough to ensure absolute convergence of the integral
but the decay is very slow which causes problems
when one tries to compute the integral \eqref{inf:Gurland} numerically.

	\item When $z>x$, catastrophic cancellations of the two terms in \eqref{2:FFF} may occur in the computation of $Y(x,s,i\tau z,-i\tau)$. In general, the computation of $Y$ using \eqref{2:FFF} is numerically stable only when $z>x$. Observe that as $\tau\to\infty$, each term of the power series of ${}_1F_2$ in \eqref{2:FFF}
has a limit, namely the series goes term-by-term to the geometric series
\[ \sum_{n=0}^\infty \left(\frac{x}{z}\right)^n .\]
This series converges only when $z>x$. The result is in agreement with \eqref{inf:Y}. Obviously, the second term on the right-hand side of \eqref{2:FFF} must be small compared with the first. However, the above is not true when $z<x$.
When we compute $Y(b)$ in \eqref{3:Phi} we can assume that $z>b$ and there is no computational issue. But when computing the integral \eqref{inf:Gurland} for $b<z<x$, we would face the issue of cancellations.
\end{enumerate}

Therefore, one should avoid numerical evaluation of \eqref{inf:Gurland} by calculating the price spread in\eqref{APspread} instead. Now we want to analyze the convergence of the integral in \eqref{1:Gurland2}
\begin{equation}\label{inf:Gurland2}
\int_0^\infty \frac1\tau\Im \Phi(b,x,s,i\tau z,-i\tau)\,d\tau.
\end{equation}
 To determine the behavior of the integrand, we use the following result from \cite[Section 10.8]{Olv}
\begin{equation}\label{inf:K}
 K_\lambda(\lambda y)\sim \left(\frac{\pi}{2\lambda}\right)^{1/2} \frac{e^{-\lambda\xi}}{(1+y^2)^{1/4}} \quad\text{as $|\arg \lambda|<\frac\pi2$, $\lambda\to\infty$},
\end{equation}
where
\[ \xi=(1+y^2)^{1/2}+\ln \frac{y}{1+(1+y^2)^{1/2}},\quad \Re y\ge 0 .\]
The values $y=0,\pm i$ have to be excluded.
We use \eqref{inf:K} with $\lambda=2\sigma^{-1} \sqrt{2iz\tau}$ and $y_1=-i\sqrt\frac{x}{z}$.
If $z<x$ we have to use the root with negative imaginary part in the calculation of $\xi_1=\xi(y_1)$.
Then we obtain
\[ F_2(x,s,i\tau z,-i\tau)\sim x^{-(1+\mu)/2}\left(\frac{\pi}{2\lambda}\right)^{1/2} \frac{e^{-\lambda\xi_1}}{(1+y_1^2)^{1/4}}\quad\text{as $\tau\to\infty$} .\]
Setting $y_2=-i\sqrt\frac{b}{z}$ and $\xi_2=\xi(y_2)$ we have
\[ F_2(b,s,i\tau z,-i\tau)\sim b^{-(1+\mu)/2}\left(\frac{\pi}{2\lambda}\right)^{1/2} \frac{e^{-\lambda\xi_2}}{(1+y_2^2)^{1/4}}\quad\text{as $\tau\to\infty$}  .\]
From \cite[Ex.\ 7.2, page 378]{Olv} we obtain in a similar way
\[ F_2'(b)\sim -b^{-\mu/2-1}\frac{\sqrt{-2i\tau}}{\sigma}\left(\frac{\pi}{2\lambda}\right)^{1/2} \frac{(1+y_2^2)^{1/4}e^{-\lambda y_2}}{y_2} .\]
This gives
\begin{eqnarray*}
 \frac{F_2(x)}{\rho F_2(b)-bF_2'(b)}&\sim & -\frac{F_2(x)}{bF_2'(b)}\\
  &\sim&\left(\frac{x}{b}\right)^{-(1+\mu)/2}\frac{\sigma}{\sqrt{-2i\tau b}} \frac{y_2}{(1+y_1^2)^{1/4}(1+y_2^2)^{1/4}} e^{-\lambda(\xi_1-\xi_2)} .
\end{eqnarray*}
Combining this with \eqref{inf:Y}, \eqref{inf:Yp}, we obtain
\begin{equation}
\Phi(b,x,s,i\tau z,-i\tau)\sim \frac{\rho}{s} \left(\frac{x}{b}\right)^{-(1+\mu)/2}\frac{\sigma}{\sqrt{-2i\tau b}} \frac{y_1}{(1+y_1^2)^{1/4}(1+y_2^2)^{1/4}} e^{-\lambda(\xi_1-\xi_2)} .
\end{equation}
It follows from this analysis that $\frac1\tau\Im \Phi(b,x,s,i\tau z,-i\tau)$ behaves like a constant times $\tau^{-3/2}e^{-a \sqrt{\tau}}$ as $\tau\to\infty$ with $\Re a>0$. This makes the integral \eqref{inf:Gurland2} numerically more attractive than the one on the right-hand side of \eqref{1:Gurland}.

\section{The limit $\sigma\to 0^+$}
It is interesting to consider the limit $\sigma\to 0^+$. We can use this limiting case to gain some insights, and also to verify theoretical and numerical results.

We take $\sigma=0$, $r<0$, $0<b<x$ (if $r>0$ then interesting results appear only for $x<b$.)
Then the geometric Brownian motion becomes deterministic
\[ X_t=x e^{rt} .\]
We define
\[ T=-\frac1r\ln \frac{x}{b}>0 .\]
so that $xe^{rT}=b$.
Then
\begin{eqnarray*}
 U_t&=&t\wedge T,\\
 V_t&=& \frac{x}{r}(e^{r(t\wedge T)}-1).
\end{eqnarray*}
The joint distribution of $(U_t,V_t)$ is a point mass which moves along a curve in the $(u,v)$-plane for $0<t<T$ and then stops at $t=T$.
The Laplace transform with respect to $(u,v)$ is
\[  \Exp^x\left[ \exp(-\alpha U_t-\beta V_t) \right]=\exp(-\alpha (t\wedge T)) \exp(-\frac{x}{r} (e^{r(t\wedge T)}-1)\beta) .\]
To shorten notation we set
\[ \gamma=-\frac{s+\alpha}{r} .\]
Then after some computation we find the triple Laplace transform \eqref{trilap} in the form \eqref{3:F} as
\begin{eqnarray*}
 \Phi_0(b,x,s,\alpha,\beta)&=&e^{\frac{\beta}{r}(x-b)}\left(\frac{b}{x}\right)^\gamma \left(\frac{1}{s}-\frac{1}{s+\alpha} M\left(1,\gamma+1,\frac{\beta b}{r}\right)\right),\\
 Y_0(x,s,\alpha,\beta)&=&\frac{1}{s+\alpha} M\left(1,\gamma+1,\frac{\beta x}{r}\right),
\end{eqnarray*}
where $M$ denotes the Kummer function.
One can confirm that $Y_0$ is the limit of the function \eqref{2:FFF} as $\sigma\to 0^+$.
In the hypergeometric series ${}_1F_2$ appearing in \eqref{2:FFF}  we can go to the limit term by term. Obviously, the second term on the right-hand side of \eqref{2:FFF} must go to zero.

Note that
\[ Z_t=\frac{V_t}{U_t}=\frac{x}{r(t\wedge T)} (e^{r(t\wedge T)}-1) .\]
Therefore,
\[ G_0(b,x,z,t)=\begin{cases} 1 & \text{if $\frac{x}{r(t\wedge T)}\left(e^{r(t\wedge T)}-1\right)\le z$},\\
0 & \text{otherwise.}
\end{cases}.\]
Suppose that
\[ \frac{x}{rT}(e^{rT}-1)<z<x .\]
Then there is a unique $t_0>0$ such that
\[ \frac{x}{rt_0}(e^{rt_0}-1)=z, \]
and we have
\[ \int_0^\infty e^{-st}G_0(b,x,z,t)\,dt= \frac{e^{-st_0}}{s} .\]
This allows us to numerically check equation \eqref{1:Gurland} in the special case $\sigma=0$.

\noindent{\bf Example:}
If $\sigma=0.1$, $r=-0.2$ then
\begin{eqnarray*}
Y(3.0,1.9,2.5.2.1)&\approx&0.094532\\
Y_0(3.0,1.9,2.5,2.1)&\approx&0.094501
\end{eqnarray*}
which is surprisingly close although $\sigma$ is not that small.
For $\Phi$ one has to make $\sigma$ smaller to get good agreement, for example, if $\sigma=0.01$, $r=-0.2$, we get
\begin{eqnarray*}
\Phi(2.0,3.0,1.9,2.5.2.1)&\approx & 1.873 \cdot 10^{-9}\\
\Phi_0(2.0,3.0,1.9,2.5,2.1)&\approx& 1.504 \cdot 10^{-9}
\end{eqnarray*}

\section{Hedging}

Delta-hedging is a common technique used by practitioners to create portfolio for managing investment risks in over-the-counter derivatives trading. The key element of this dynamic hedging strategy is to trade the underlying asset (or futures on the assets) according to the sensitivity measure, known as the delta,
\[\Delta_{\mathrm{AP}_b}=\frac{\partial}{\partial x} \left( \mathrm{AP}_b\right).\]
 As analytical computation methods of the conditional Asian option is not previously known in the literature,  BNP Paribas' calculation of the delta of the conditional Asian option is likely carried out by Monte Carlo simulations. Bear in mind that conditional Asian options are long-dated options with a typical term of 5 years. A foreseeable difficulty with such an approach is that high efficiency of the conditional Asian option price itself can only be achieved at the cost of very expensive computations. Since the derivatives are approximated by difference quotients, the computational burden is exacerbated with repeated calculations of option prices with small changes in parameters. A small sampling error in the estimation of option prices can lead to a large relative error in estimating sensitivity measures such as deltas. It may happen that the inevitable sampling errors of the estimation statistics overrun subtle differences between option prices resulted from small changes in parameters, in which case the estimation of the delta is no longer reliable. Therefore, analytical methods can be of vital importance in the computation of greeks. Although we only compute the delta in this paper, the methodology can be easily extended to other greeks.

Owing to the computational advantage of the price spread, we also break down the delta into two parts in the same way as in \eqref{APspread}.
\[\Delta_{\mathrm{AP}_b}=\Delta_{\mathrm{AP}_0}+ e^{-rT}\int^K_0 \frac{\partial}{\partial x} D(b,x,z,T) \df z .\]
The two quantities can be computed as follows.
\[\Delta_{\mathrm{AP}_0}=\frac1Te^{-rT}\frac{\partial}{\partial x} Q(x,T, TK), \]
where $(\partial /\partial x) Q$ can be determined numerically by its Laplace transform
\[\frac{\partial}{\partial x} \tilde{Q}(x,s,y)=\frac{16}{\sigma^4} \tilde{Q}_0 \left( \frac{4s}{\sigma^2},\frac{\sigma^2 y}{4x}\right)-\left( \frac{y}{x}\right) \tilde{P}\left( \frac{4s}{\sigma^2},\frac{\sigma^2 y}{4x}\right).\]
Similarly, one can show that $(\partial /\partial x) D$ can be determined numerically by its Laplace transform
\be \int^\infty_0 e^{-st} \frac{\partial}{\partial x}  D(b,x,z,t) \df t=\frac1{\pi} \int^\infty_0 \frac1{\tau} \Im \frac{\partial}{\partial x} \Phi(b,x,s,i\tau z, -i\tau) \df \tau, \label{Dprime}\ee
where
\by
&&\frac{\partial}{\partial x} \Phi(b,x,s,i\tau z, -i\tau)=\frac{\rho\left(\frac1s-Y(b)\right)+bY'(b)}{\rho F_2(b)-bF_2'(b)} F'_2(x);\\
&&F'_2(x)=\frac{\lambda-\mu-1}2 x^{-(3+\mu)/2} K_{\lambda}\left( \frac2\sigma \sqrt{2\beta x}\right)-\frac{\sqrt{2\beta}}{\sigma} x^{-(2+\mu)/2} K_{\lambda+1}\left( \frac2\sigma \sqrt{2\beta x}\right).
\ey
Even though \eqref{Dprime} works for all $z$, we can avoid the numerical integration for $0\le z\le b$ by using
\[ \frac{\partial}{\partial x}  D(b,x,z,T) =\frac{\partial}{\partial x} G(0,x,z,T) =\frac{\partial}{\partial x} P(x,T,Tz), \]
where $(\partial/\partial x) P$ can be determined numerically by its Laplace transform \[\frac{\partial}{\partial x}\tilde{P}(x,s,y)=-\left(\frac{y}{x^2}\right) \tilde{p}_0\left( \frac{4s}{\sigma^2}, \frac{\sigma^2y}{4x} \right)\] with $\tilde{p}_0$ given by the explicit expression \eqref{p0}.

\section{Numerical examples} \label{sec:num}

The computation of regular Asian option price is well-known and extensively studied in the literature. Nevertheless, we use the first example to confirm the accuracy of the numerical algorithm for regular Asian put option price in Section \ref{sec:asian}, which serves as part of the numerical algorithm for conditional Asian put option price in the second example. Here we use the set of parameters under which Asian option prices have been computed in many papers by a variety of techniques including Monte Carlo simulations in \cite[Table 4, Row 2]{FuMadWan}, inverse Laplace transform in \cite[Test Problem 2]{Sha}, numerical PDE method in \cite{Vec} and eigenfunction expansion in \cite[Case \#5]{Lin04b}, etc. Similar comparisons on many more numerical techniques can be seen in \cite{FuMadWan} and \cite{Lin04b}.
\[r=0.05, \quad \sigma=0.50, \quad T=1, \quad x=2.0,\quad K=2.0.\]
It should be noted that nearly all works regarding Asian options in the literature are done on call options. Here we use the formulas in Section \ref{sec:asian} to compute the Asian put option price with the set same of parameters first and determine the call option price by the put-call parity.
\begin{table}[h] \centering
\begin{tabular}{|c|c|c|c|c|}
  \hline
  \cite{FuMadWan} & \cite{Sha} & \cite{Vec} & \cite{Lin04b} & Section \ref{sec:asian}  \\ \hline
  $0.249$(MC100) & $0.246416$ & $0.246$ & $ 0.2464156905$ & $0.2464156819$ \\ \hline
\end{tabular}
\caption{Verification of regular Asian option prices} \label{tab:asian}
\end{table}

In this second example, we consider the pricing and hedging of long-term at-the-money conditional Asian options. Besides demonstrating this new approach of conditional Asian option pricing, we also intend to put to the test BNP Paribas' cost/benefit claim that conditional Asian options can achieve $75\%$ of initial delta of regular Asian options with a reduction of $40\%$ in costs. We shall use the following set of common parameters
\[ \quad r=0.05, \quad b=1.0,\quad T=5,\quad x=2.0,\quad  K=2.0 .\]
The computation of conditional Asian put option price is carried out in the following five-step procedure.
	\begin{enumerate}
	\item[Step 1:($\hat{D}$)] We calculate the integrals for $s=(i \ln 2)/T, i=1,2, \cdots, 2M,$
\begin{equation}\label{num:int}
 \int_0^\infty \frac1\tau\Im\Phi(b,x,s,i\tau z,-i\tau)\,d\tau
\end{equation}
appearing on the right-hand side of \eqref{1:Gurland2} for $z=b+hk$ and $k=1,2,\cdots,(K-b)/h$.
 The integrals are computed by first determining a number $a>0$ so that the truncation error of \eqref{num:int} is less than $10^{-2.2M}$,
\begin{equation}\label{num:int2}
\hat{D}(b,x,z,s)= \int_0^a \frac1\tau\Im\Phi(b,x,s,i\tau z,-i\tau)\,d\tau .
\end{equation}

	\item[Step 2:($D$)]  The values of $D$ are computed by the Gaver-Stehfest algorithm
\[D(b,x,z,T)=\frac{\ln(2)}T \sum^{2M}_{k=1} \xi_k \hat{D}\left(b,x,z, \frac{k\ln(2)}T  \right)\]
where
\[\xi_k=(-1)^{M+k} \sum^{k\wedge M}_{j=\lfloor (k+1)/2\rfloor} \frac{j^{M+1}}{M!} {M \choose j} {2j \choose j}{j \choose k-j},\] with $\lfloor x \rfloor$ being the integer part of $x$ and $k\wedge M=\min\{k,M\}$.
	\item[Step 3: ($\mathrm{AP}_0$)] The regular Asian put option price is determined by the inverse Laplace transform method outlined in \eqref{ap0}, \eqref{eq:Q} and \eqref{eq:Q0}.
	\item[Step 4: (Spread)] $D(b,x,z,T)$ has been determined in Step 1 for $z\ge b$. $D(b,x,z,T)=G(0,x,z,T)$ when $z\le b$, which can be determined by \eqref{GP}. The quantity $Q$ is then computed numerically by inverting its Laplace transform \eqref{eq:P} and \eqref{eq:tP0}. Then the integral in \eqref{APspread} is computed last by the trapezoidal rule.
	\item[Step 5:($\mathrm{AP}_b$)] The final result is produced by \eqref{APspread}.
\end{enumerate}

Numerical algorithms in this paper are implemented in Mathematica on a Macbook with 2.9 GHz Intel Core i5 CPU. The Gaver-Stehfest algorithm for numerical inversion of the Laplace transform utilizes $M$ terms in the Salzer sequence acceleration scheme. It is stated in \cite{AbaWhi06} that the algorithm requires a system precision of $2.2 M$ and produces roughly $0.90M$ significant digits. In our numerical examples, we use $M=5$ and the system precision of $11$ digits in Mathematica, which is expected to produce about 4 or 5 correct digits for $D(b,x,z,t)$. The integrals \eqref{num:int2} are computed with Mathematica's numerical integration routine without any specification. In Steps 3 and 4, $G(0,x,z,t)$ and $\int_0^z G(0,x,w,t)\df w$ are computed using the Laplace transform method introduced in Section \ref{sec:asian}. The needed inverse Laplace transforms are computed with the Euler algorithm in \cite{AbaWhi06}.  In Step 5, we use the trapezoidal rule with step size of $h=0.1$ for the numerical integration in \eqref{APspread}.

It should be pointed out that only the first step in the five-step procedure, the computation of the integrals \eqref{num:int2}, is time consuming whereas all other steps can be done within a second. The computation time for the 110 integrals is about 20 minutes.
The computation time is longest for $z$ between $b$ and $x$ but much shorter when $z>x$. The reason is that the integrand decays faster as $\tau\to\infty$ when
$z>x$. This is confirmed by the asymptotics we derived in Section \ref{sec:asyminfty}.

\begin{figure}[h]
\begin{center}
\includegraphics[width=0.4\textwidth]{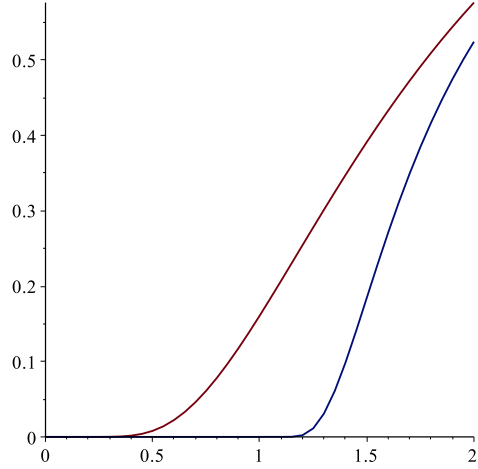}
\end{center}
\caption{Distributions of $Z_T$: $G(0,2,z,5)$ (top) and $G(1,2,z,5)$ (bottom).}
\label{num:fig}
\end{figure}

Figure \ref{num:fig} shows the distribution functions of the average-to-date at maturity. The top red line represents $z\to \Pro(Y_T/T\le z)$ where $Y_T/T$ is the (unconditional) average-to-date of asset prices with $Y$ defined in \eqref{Y}. The bottom blue line depicts $z\to \Pro(Z_T \le z)$ where $Z$ defined in \eqref{Z} is the average-to-date above the observation threshold $b$. The graph provides some geometrical interpretation of the option prices. The areas underneath the red line and the blue line between $z=0$ and $z=K$ represent the accumulated value at maturity of regular Asian put premium and conditional Asian put premium ($e^{rT}\mathrm{AP}_0$ and $e^{rT}\mathrm{AP}_b$) respectively. The area below the red line and above the blue line accounts for the accumulated value of the price spread between the two options.

We make a comparison of regular and conditional Asian put options in Table \ref{tab:comp} under various assumptions of volatility coefficient $\sigma$, which suggests very dramatic changes in the cost/benefit comparison of the two types of options. The results are generally not as sensitive to the risk-free rate $r$.  While the highlighted case of $\sigma=0.4$ does support the claim that the conditional Asian put costs $60\%$ of its counterpart Asian put and achieves a delta $75\%$ of that of Asian put, such a result should not be over-generalized. This result appears to indicate that the simulations by BNP Paribas may have been performed with a volatility at around $40\%$. 

In general, the difference on per unit cost of delta between a conditional Asian option and classical Asian option  is much pronounced with high volatility than that with low volatility. For example, when $\sigma=0.6$, the per unit cost of delta for conditional Asian option is about 0.8675 whereas that for Asian option is about 1.4389. When $\sigma=0.2$, the per unit cost of delta for conditional Asian option is about 0.3485 whereas that for Asian option is about 0.3580. An intuitive explanation can be given as follows. When the volatility is low, it is less likely that asset prices drop below the threshold and hence there is very little difference in average prices between the two options. When the volatility is high, it is more likely that asset prices below the threshold are filtered out by the average price in conditional Asian option and hence the difference between the two options becomes more drastic.

\begin{table}[h] \centering
\begin{tabular}{|c|c|c|c|c|c|c|}
  \hline
  $\sigma$ & $\mathrm{AP}_b$ & $\mathrm{AP}_0$ & $\mathrm{AP}_b/\mathrm{AP}_0$ &  $\Delta_{\mathrm{AP}_b}$  &$\Delta_{\mathrm{AP}_0}$ & $\Delta_{\mathrm{AP}_b}/\Delta_{\mathrm{AP}_0}$  \\ \hline
  $0.6$ & $0.1669$ & $0.4026$  & $41.47\%$  & $-0.1924$ & $-0.2798$ & $68.76\%$ \\ \hline
  $0.5$ & $0.1625$ & $0.3256$ & $49.92\%$ & $-0.2029$ & $-0.2859$  & $70.97\%$ \\ \hline
   \marktopleft{c2} $0.4$ & $0.1530$ & $0.2465$ & $62.08\%$ & $-0.2156$ & $-0.2871$ &  $75.11\%$ \markbottomright{c2}\\ \hline
  $0.3$ & $0.1295$ & $0.1664$ & $77.86\%$ & $-0.2316$ & $-0.2782$  & $83.27\%$ \\ \hline
 $0.2$ & $0.0810$ & $0.0877$ & $92.42\%$ & $-0.2324$ & $-0.2450$  & $94.84\%$\\ \hline
\end{tabular}
\caption{Comparison of regular and conditional Asian options} \label{tab:comp}
\end{table}

\section*{Appendix - Lommel's functions}

The inhomogeneous linear differential equation
\[z^2 y''+z y'+(z^2-\lambda^2) y=k z^{\mu+1}\] has two solutions $k s_{\mu,\lambda}(z)$ and $k S_{\mu,\lambda}(z)$ known as Lommel's functions, where
\by
s_{\mu,\lambda}(z)&=&\frac{z^{\mu+1}}{(\mu-\lambda+1)(\mu+\lambda+1)}\,_1F_2\left(1;\tfrac12(\mu-\lambda+3),\tfrac12(\mu+\lambda+3);-\frac14 z^2\right);\\
S_{\mu,\lambda}(z)&=&s_{\mu,\lambda}(z)\\
&&+\frac{2^{\mu-1}}{\sin(\lambda \pi)}\Gamma\left(\tfrac12 (\mu-\lambda+1)\right) \Gamma\left( \tfrac12(\mu+\lambda+1)\right)\left[\cos\left( \tfrac12(\mu-\lambda)\pi\right) J_{-\lambda}(z)-\cos \left(\tfrac12 (\mu+\lambda)\pi\right) J_\lambda(z)\right],
\ey
and $J_\nu$ is a Bessel function; see \cite[Section 10.7]{Wat}. It is known from \cite[Section 10.75]{Wat} that 
\be S_{\mu,\lambda}(z)\sim z^{\mu-1}\quad\text{as $z\to\infty$}.\label{Sasymp}\ee
 We define a modified Lommel's function
\[T_{\mu,\lambda}(z)=e^{-i \pi(\mu+1)/2} s_{\mu,\lambda}(iz)-2^{\mu-1}\Gamma\left(\tfrac12 (\mu-\lambda+1)\right) \Gamma\left( \tfrac12(\mu+\lambda+1)\right)I_\lambda(z).\]
 Then it can be shown that $kT_{\mu,\lambda}(z)$ is a solution to the differential equation
\[z^2 y''+z y'-(z^2+\lambda^2) y=k z^{\mu+1}.\] The function $T$ is real-valued if both $\mu, \lambda \in \R$ and $z>0$.
Observe that the function can also be represented in terms of $S_{\mu,\lambda}$
\[T_{\mu,\lambda}(z)=e^{-i \pi(\mu+1)/2} S_{\mu,\lambda}(iz)-\frac{2^{\mu}}{\pi}e^{-i(\lambda+\mu+1)\pi/2} \cos \left(\tfrac12(\mu-\lambda)\pi\right)\Gamma\left(\tfrac12 (\mu-\lambda+1)\right) \Gamma\left( \tfrac12(\mu+\lambda+1)\right)K_\lambda(z).\]
It follows from \eqref{Sasymp} that
\be T_{\mu,\lambda}(z)\sim -z^{\mu-1}\quad\text{as $z\to\infty$}.\label{Tasymp}\ee
 Therefore, the general solution to \eqref{2:ode}  is given by
\[F(x)=C_1 F_1(x)+C_2 F_2(x)+k x^{-(\mu+1)/2} T_{\mu,\lambda}\left(\frac{2\sqrt{2\beta x}}{\sigma}\right),  \]
where $C_1$ and $C_2$ are arbitrary constants, and
\[k=-\frac{2}{\sigma^2} \left(\frac{\sqrt{2\beta}}{\sigma}  \right)^{-1-\mu}.\] It follows from the properties of $F_1, F_2$ and \eqref{Tasymp} that the solution to \eqref{2:ode} satisfying both boundary conditions \eqref{2:bc1} and \eqref{2:bc2} must be $F(x)=k x^{-(\mu+1)/2} T_{\mu,\lambda}(2\sqrt{2\beta x}/\sigma)$ which can be rewritten as \eqref{2:FFF}.

\bibliography{central-bibliography}
\end{document}